\documentclass[12pt]{article}
\usepackage{epsfig}
\begin{document}
\begin{center}
{\Large{
The Chandrasekhar's Equation for Two-Dimensional Hypothetical White Dwarfs}}

\bigskip
{\large{
Sanchari De and Somenath Chakrabarty$^\dagger$

\medskip
Department of Physics, Visva-Bharati, Santiniketan-731235,
India\\
$^\dagger$somenath.chakrabarty@visva-bharati.ac.in}}

\medskip
Pacs:73.20.-r, 71.10.Ca, 21.65.+f, 13.75.Cs

\end{center}
\newpage
\begin{center}
Abstract
\end{center}

In this article we have extended the original work of Chandrasekhar on
the structure of white dwarfs to the two-dimensional case. Although
such two-dimensional stellar objects are hypothetical in nature, we
strongly believe that the work presented in
this article has some academic interest. In particular it is a 2-D version of Newtonian gravity. The electromagnetic or
the coulomb problem in two dimension has already been investigated in greater detail, however, this particular problem
with the same logarithmic type potential has not been studied before.

\section{introduction}
White dwarfs are the end product of intermediate massive and low mass stars. 
At the late stage of evolution of such stars, because of intense
stellar wind, there are huge loss of  
matter from the outer part of such stellar objects. This is some kind of
instability in this type of stars developed at the late stage
of evolution. Most of the mass of these 
stars are thrown out and they form a 
gaseous nebula like structure, which can be observed by high power
optical telescope.
The compact object at the central region of this gaseous nebula is
called the white dwarf. The mass of this compact object is
very close to the solar mass with the size of a big planet (Jupiter like). 
Because of this very reason the gaseous region along with the massive central object is called 
planetary nebula. The formation of planetary nebula is a quite process 
compared to supernova explosion of massive
stars and super-massive stars, which produce neutron star or in the
extreme case, black holes \cite{ST,TP,ARC}.

Now depending on whether the mass of the progenitor  is 
comparable with the solar mass or heavier than sun,
inside the main-sequence stars the conversion of 
hydrogen to helium is going on continuously through either p-p
chain reactions or CNO cycles \cite{TP,ARC,RQ}.
To study the structure of such main-sequence stars a polytropic equation of 
state of the matter is considered. These stars
are sometime also called main-sequence polytropic stars. The form of
polytropic equation of state is given by \cite{ST,TP,ARC,RQ} 
\begin{equation}
P=K\rho^{\Gamma}
\end{equation}
where $P$ is the kinetic pressure of the constituents, $\rho$ is the mass 
density, $K$ and $\Gamma$ are constants.
Both these constants depend on the physical properties of the matter. 
The constant $\Gamma$ is called the polytropic index, which is more or 
less like the adiabatic index $\gamma$. Now the stellar matter inside the 
main-sequence stars are mainly a fully ionized hydrogen plasma. 
Further, the temperature of the matter is so high that the plasma is non-degenerate 
in nature. The temperature is also high enough for the hydrogen ions 
to overcome the inter-ionic coulomb barrier and undergo thermo-nuclear 
fusion reaction to form helium. The gravitational collapse
of these stars are opposed by the kinetic pressure of the matter. To
study the structure and gross properties of these
main-sequence stars, the well known Lane-Emden equation is used \cite{ST,TP,ARC,RQ}. The 
Lane-Emden equation is essentially obtained by combining 
the hydrostatic equilibrium equation with the polytorpic
equation of state of the stellar matter. 

On the other hand there is no thermo-nuclear reactions inside the white 
dwarf stars. The object is mainly made up of dense and fully ionized
carbon or oxygen matter at the inner region with a crust of helium and a 
very thin layer of hydrogen gas at the skin of the star. 
The crustal matter is also fully ionized because of very high matter 
density. In this case the stability against gravitational collapse is 
governed by the degeneracy pressure of electron gas. Whereas the mass
of the white dwarf star comes from the baryons, which are assumed to
be at rest.

Therefore in general the equation of state of matter inside a white dwarf can not be 
represented by the polytropic form. In this case not the kinetic pressure, 
but the degeneracy pressure of the electron gas will contribute in the 
hydrostatic equilibrium condition. The form of the equation to study the 
structure and gross properties of white dwarf stars was originally derived by
Chandrasekhar \cite{RQ,CH}, which coincides with the Lane-Emden equation only in some
special situation.

In this article we have made an extension of Chandrasekhar's original work and
investigated the gross properties of two dimensional white dwarf stars. 
Since there is no physical existence of such objects, therefore just for 
the sake of academic interest we have developed this formalism \cite{RQ,CH}.
To the best of our knowledge, this problem has not been reported before.

\section{Basic Formalism}
The surface density of degenerate electron gas is given by \cite{LL,HU}
\begin{equation}
n_e=\frac{N}{S}=\frac{p_F}{2\pi\hbar^2}
\end{equation}
where $p_F$ is the electron Fermi momentum. Hence the mass density
\begin{equation}
\rho=n_e\mu_em_p
\end{equation}
where $m_p$ is the baryon mass  and $\mu_e$ is the electron mean 
molecular weight. It can very easily be shown that the expression for 
pressure of the degenerate electron gas in two-dimension is given by 
\begin{equation}
P=\frac{c^2}{2\pi\hbar^2}\int_0^{p_F}\frac{p^3dp}{(p^2c^2+m_e^2c^4)^{1/2}}
\end{equation}
where $m_e$ is the electron rest mass. Defining $y=p/m_ec$ and 
$\xi =p_F/m_ec$, we have
\begin{equation}
P=Cf(\xi)
\end{equation}
where 
\begin{equation}
C=\frac{(m_ec^2)^3}{2\pi(\hbar c)^2}
\end{equation}
and
\begin{equation}
f(\xi)=\int_0^{\xi}y^2d(1+y^2)^{1/2}
\end{equation}
similarly we have for the mass density
\begin{equation}
\rho=C^\prime\xi^2
\end{equation}
where
\begin{equation}
C^\prime=\mu_e\frac{m_p(m_ec^2)^2}{2\pi(\hbar c)^2}
\end{equation}
Now the hydrostatic equilibrium equation for white dwarf stars is given by \cite{ST,RQ}
\begin{equation}
\frac{dP}{dr}=-g(r)\rho(r)
\end{equation}
Now in two-dimension
\begin{equation}
g(r)=\frac{G}{r}\int_0^r 2\pi r^\prime\rho(r^\prime)dr^\prime
\end{equation}
Substituting $g(r)$ into eqn.(10) and then differentiating with respect to 
$r$, then  after rearranging some of the terms, we have
\begin{equation}
\frac{1}{r}\frac{d}{dr}\left (\frac{r}{\rho}\frac{dP}{dr}\right )+2\pi G\rho=0
\end{equation}
One can start from this equation to obtain Lane-Emden equation in two 
dimension using polytropic equation of state.
However, in this article we shall not discuss Lane-Emden equation. In
some other communication we shall investigate the
structure and gross properties of two-dimensional polytropic
main-sequence stars using the Lane-Emden equation \cite{SS}.
Now substituting $P$ and $\rho$ from eqn.(5) and eqn.(8) in
eqn.(12), we get
\begin{equation}
\frac{C}{C^\prime}\frac{1}{r}\frac{d}{dr}\left (\frac{r}{\xi}\frac{df}{dr}
\right )+2\pi G C^\prime\xi^2=0
\end{equation}
Since
\begin{equation}
\frac{1}{\xi^2}\frac{df}{dr}=\frac{d}{dr}(1+\xi^2)^{1/2}
\end{equation}
we have after substituting $x^2=1+\xi^2$
\begin{equation}
\frac{1}{r}\frac{d}{dr}\left (r\frac{dx}{dr}\right )+
\frac{2\pi GC^{\prime^2}}{C}(x^2-1)=0
\end{equation}
Therefore $x=1$ for $\xi=0$, the extreme non-relativistic situation, 
whereas $x\rightarrow\infty$, the ultra-relativistic condition.
Then Following \cite{RQ}, we substitute
\[
U=\frac{x}{x_c} ~~{\rm{and}}~~ Z=\frac{r}{A}
\]
where $x_c$ represents the value of $x$ at the centre and $A$ is an
unknown constant. Therefore $U=1$ at the centre. Then we have
\begin{equation}
\frac{1}{Z}\frac{d}{dZ}\left (Z\frac{dU}{dZ}\right )+
\frac{2\pi GC^{\prime^2}x_c}{C}A^2\left (U^2-\frac{1}{x_c^2}\right )=0
\end{equation}
Choosing 
\begin{equation}
A=\left (\frac{C}{2\pi Gx_c}\right )^{1/2}\times\frac{1}{C^{\prime}}
\end{equation}
we have
\begin{equation}
\frac{1}{Z}\frac{d}{dZ}\left (Z\frac{dU}{dZ}\right )+(U^2-\frac{1}{x_c^2})=0
\end{equation}
This is the differential equation describing the structure of white dwarfs 
in two-dimension, i.e., it is the modified version of Chandrasekhar 
equation. The corresponding three dimensional form, which was first 
obtained by Chandrasekhar, is given by  \cite{RQ,CH}
\begin{equation}
\frac{1}{Z^2}\frac{d}{dZ}\left (Z^2\frac{dU}{dZ}\right )+\left (U^2-\frac{1}{x_c^2}\right )^{3/2}=0
\end{equation}
Just like the original version of Chandrasekhar's  equation  (eqn.(19)), the 
differential equation, given by eqn.(18) can not be solved analytically. 
To obtain the numerical solution for this second
order differential equation, we use the following initial conditions
At the centre, $r=0$, i.e., $Z=0$ and $U=1$, which is the maximum value of 
$U$, therefore at the centre $dU/dZ=0$. On the other hand, the surface of 
the white dwarf is obtained from the following condition. At
$Z=Z_s$, the surface value, $\rho=0$, therefore $\xi_s=0$, $x_s=1$ and
$U_s=1/x_c$. Then from eqn.(8), we can write down the expression for matter
density in the following form
\begin{equation}
\rho=C^{\prime}x_c^2\left (U^2-\frac{1}{x_c^2}\right )
\end{equation}
The radius of the  white dwarf star is given by
\begin{equation}
R=AZ_s=\left (\frac{C}{2\pi Gx_c}\right )^{1/2}\times \frac{Z_s}{C^{\prime}}
\end{equation}
and the corresponding mass of the white dwarf can be obtained from the 
integral
\begin{equation}
M=2\pi\int_0^Rrdr\rho(r)
\end{equation}
Using $\rho(r)$ from eqn.(20) and changing the integration variable to 
$Z$, we have 
\begin{equation}
M=-\frac{Cx_cZ_s}{GC^{\prime}}\frac{dU}{dZ}\mid_{Z=Z_s}
\end{equation}
In the evaluation of the mass $M$ of the white dwarf, we use $\mu_e=2$, 
assuming that there is no hydrogen in white dwarf
star and the surface value of the gradient $dU/dZ$ is obtained from the 
numerical solution of eqn.(18).

In fig.(1) we have shown the variation of mass of the white dwarf
stars with $x_c$, or indirectly with the central density of the star.
Whereas in fig.(2) the variation of radius of the star with $x_c$ is
shown. From the figures it is quite obvious that a white dwarf
star in two-dimension becomes more compact in size but at the same time massive with the
increase in central density. The mass of the star becomes infinitely
large as the central density tending to infinity. In fig.(3) we have
shown the mass-radius relation for such objects.
We have chosen the upper limit of $x_c$ in such a way that the mass of the object
become $\approx 1.41M_\odot$ \cite{HS,CG}. 
\section{Conclusion}
In this article we have extended the original idea of Chandrasekhar. The formalism developed is for the two-dimensional
white dwarf stars. Of course the objects are hypothetical in nature. Although in 2-D Newtonian gravity the potential
has the same logarithmic nature as in the case of Coulomb problem, on which a lot of work has been done, there was no
reported result on 2-D version of Chandrasekhar equation.
\begin{figure}[ht]
\psfig{figure=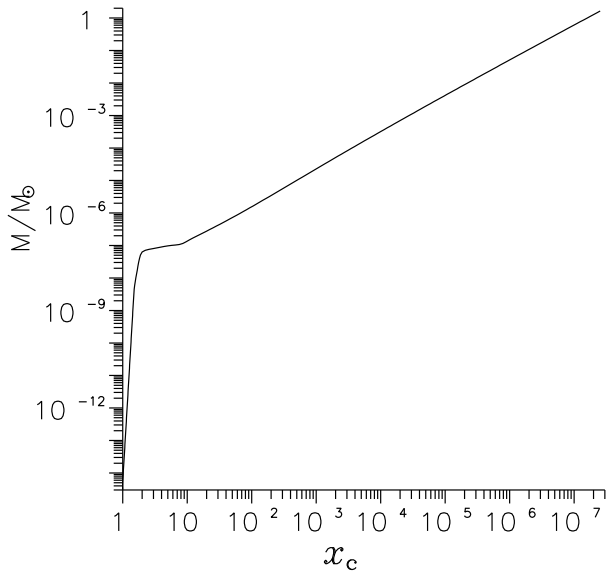,height=0.5\linewidth}
\caption{Variation of mass with $x_c$}
\end{figure}
\begin{figure}[ht]
\psfig{figure=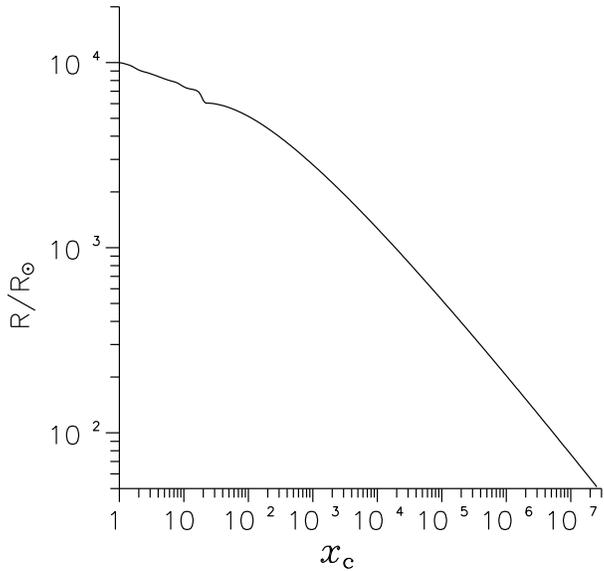,height=0.5\linewidth}
\caption{Variation of Radius with $x_c$}
\end{figure}
\begin{figure}[ht]
\psfig{figure=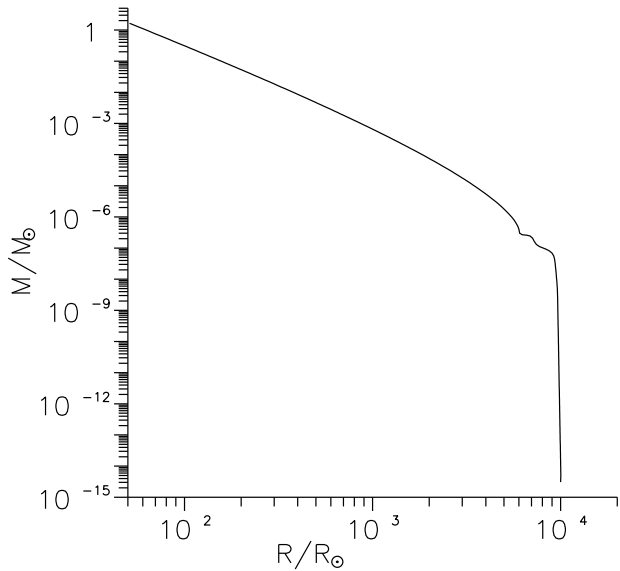,height=0.5\linewidth}
\caption{Mass-Radius relation}
\end{figure}


\begin{thebibliography}{}
\bibitem{ST} S.L. Shapiro and S.A. Teukolsky, Black Holes, White Dwarfs
and Neutron Stars, John Wiley and Sons, New York, (1983).
\bibitem{TP} T. Padmanabhan, Theoretical Astrophysics, Vol. II: Star and Stellar Systems, Cambridge University
Press, 2001.
\bibitem{ARC} A. Rai Choudhuri, Astrophysics for Physicists, Cambridge University
Press, 2010.
\bibitem{RQ} H.Q. Huang and K,N, Yu, Stellar Astrophysics, Springer,
(1998).
\bibitem{CH} S. Chandrasekhar, An Introduction to the Study of Stellar Structure, Univ. of Chicago Press, 1939..
\bibitem{LL} L.D. Landau and E.M. Lifshitz, Statistical Physics, Part-I, Butterworth-Heimenann, Oxford, 1980.
\bibitem{HU} K. Huang, Statistical Mechanics, Wiley-Eastern Pvt. Ltd., New Delhi, 1975.
\bibitem{SS} Sanchari De and Somenath Chakrabarty, to be communicated.
\bibitem{HS} T. Hamada and E.E. Salpeter, Astrophys. Jour., {\bf{134}}, 683 (1961).
\bibitem{CG} J.P. Cox and R.T. Giuli, Stellar Structure, Gordan and Breach Science Publishers Inc, 1968.
\end{thebibliography}
\end{document}